\documentstyle[12pt,psfig]{article}
\begin{document}

\vskip 1truecm
\rightline{Preprint  MCGILL-97/36}
\rightline{ e-Print Archive: hep-ph/9801204}
\vspace{0.2in}
\centerline{\Large A Nonperturbative Measurement }
\vspace{0.15in}
\centerline{\Large of the Broken Phase Sphaleron Rate} 
\vspace{0.3in}

\centerline{\large Guy D. 
	Moore\footnote{e-mail: guymoore@physics.mcgill.ca}}

\medskip

\centerline{\it Dept. of Physics, McGill University}
\centerline{\it 3600 University St.}
\centerline{\it Montreal, PQ H3A 2T8 Canada}

\medskip

\centerline{\bf Abstract}

We develop a method to compute the sphaleron rate in the electroweak 
broken phase nonperturbatively.  The rate is somewhat slower than a 
perturbative estimate.  In SU(2)$\times$U(1) Higgs theory at the physical
value of $\Theta_W$, and assuming that the latent heat of the phase
transition reheats the universe to the equilibrium temperature,
baryon number erasure after the phase transition is
prevented only when $(\lambda/g^2) \le 0.036$.

\smallskip

\begin{verse} 
PACS numbers:  11.15.Ha, 11.15.Kc
\end{verse}

\begin{verse}
Key words:  sphaleron, baryogenesis, baryon number violation,
electroweak phase transition, lattice gauge theory
\end{verse}

\section{Introduction}

In the past ten years the effort to understand electroweak baryogenesis
has driven the study of the physics of the hot electroweak plasma at and
around the electroweak phase transition.  The relevant physics has been
put on a much sounder foundation, and in particular we now possess
reliable nonperturbative studies of the strength of the phase transition
and of its disappearance at sufficiently high scalar (Higgs)
self-coupling \cite{KLRSresults,KLRSresult2,Kripfganz}.

Electroweak baryogenesis in the minimal standard model is now
ruled out on two independent grounds.  First, it cannot produce enough
baryons because there is not enough CP violation in the minimal standard
model \cite{Gavela}.  Second, even if it 
could produce the baryons, they would be
wiped out in the aftermath of the phase transition, because the rate of
baryon number violating processes does not shut off fast enough after
the phase transition.  This is because it is excluded that the standard
model Higgs boson can have a weak enough self-coupling to supply the
required strength to the phase transition \cite{KLRSresults}.
In fact, experimental bounds on the Higgs mass are now high enough that
if the minimal standard model is the right physical theory, we can
conclude that there was {\em no} cosmological electroweak phase transition.

Extensions of the standard model, such as the minimal supersymmetric
standard model with a light stop, appear to be alive and kicking for at
least part of their parameter space.  To determine what parts of their
parameter space are viable, we demand that they pass two tests.  There
must be enough CP violation to produce sufficient baryon number to
explain astrophysical observations; and baryon number must not be
subsequently erased.  The production depends on where CP violation appears
in the theory and on the mechanism by which it causes a baryon number
surplus.  We will not address this question here.  
The erasure depends on the efficiency of baryon number violating
processes {\it after} the phase transition.  
This paper will study this erasure nonperturbatively.

\begin{figure}[t]
\centerline{\mbox{\psfig{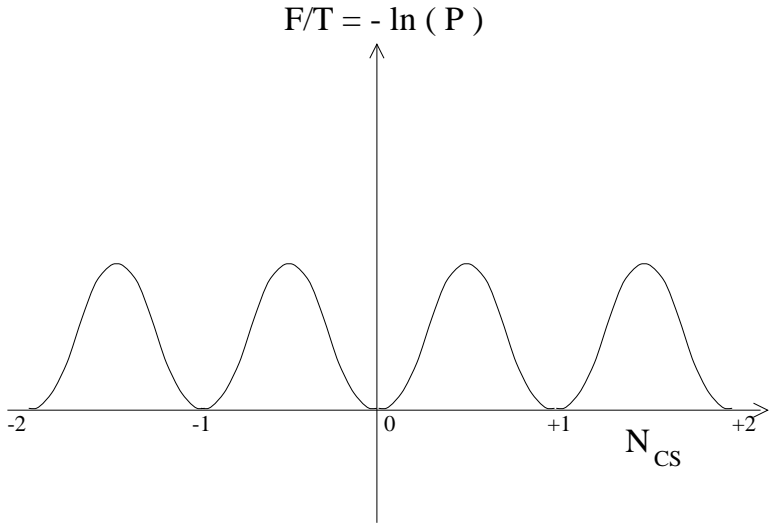}}}
\caption{\label{Fig1}  ``Cartoon'' of the free energy dependence on
$N_{\rm CS}$.}
\end{figure}

Arnold and McLerran established the basic picture for baryon number
violation in the electroweak broken phase \cite{ArnoldMcLerran}.  
Baryon number change in the
electroweak theory is proportional to change in the Chern-Simons number
($N_{\rm CS}$) of the SU(2) fields, because of the anomaly
\cite{tHooft}.  But in the broken phase, there is a barrier to changing
$N_{\rm CS}$.  To change from one integer value to another, one must
pass through a half integer state, and while there are configurations
of integer $N_{\rm CS}$ with arbitrarily small energy, the energy of
a configuration of half integer $N_{\rm CS}$ is bounded by the energy
of Klinkhamer and Manton's ``sphaleron'' configuration \cite{Manton}.
At finite temperature the same statements are true for free energy, and
one can use Langer's formalism \cite{Langer,Affleck}
to find the rate of baryon number violation.  The classic cartoon of
free energy as a function of $N_{\rm CS}$ is shown in Fig. \ref{Fig1}.  The
hard part of getting from, say, $N_{\rm CS} = 0$ 
to $N_{\rm CS} = 1$ is getting over
the peak (the sphaleron barrier) in between.  Let us assume
that, whenever the system passes over the peak, it subsequently gets
stuck in the new minimum.  Then the diffusion constant for 
$N_{\rm CS}$ is set by the flux of states in the thermal ensemble 
going over the barrier.  The diffusion constant for $N_{\rm CS}$,
\begin{equation}
\gamma_d \equiv \lim_{t \rightarrow \infty} \frac{
	\langle ( N_{\rm CS}(t) - N_{\rm CS}(0) )^2 \rangle}{t} \, ,
\end{equation}
is the probability per unit time of crossing the peak, which is the 
flux of states in the thermal ensemble over the peak.  The flux of 
states over the peak is the probability to have $N_{\rm CS}$ 
within $\epsilon / 2$ of the peak, times the
mean value of $ | dN_{\rm CS} / dt | $ measured at the peak, divided 
by $\epsilon$.  Since the
diffusion constant should be extensive, we generally refer to the
diffusion constant per unit volume, $\Gamma_d \equiv \gamma_d / V$,
often called the sphaleron rate.

The height of the sphaleron
barrier in the immediate aftermath of the electroweak
phase transition depends on the broken phase
expectation value of the Higgs condensate, $\phi \equiv \sqrt{ \langle 
\phi^2 \rangle}$, at the ambient temperature.  In the minimal
standard model the latent heat of the transition is not quite enough to 
heat the plasma back to the equilibrium temperature, but in
supersymmetric extensions, it may be; so we need the jump in
$\langle \phi^2 \rangle$ at $T_{\rm c}$.
We now have reliable nonperturbative studies of this quantity
\cite{KLRSresults,Kripfganz}.
These studies were motivated largely so
the determined value of $ \langle \phi^2 \rangle$ 
could be plugged into the
calculation of the sphaleron rate.  However, the present state of the
art in the sphaleron rate is a one loop calculation.  At the equilibrium
temperature, we know that perturbation theory for infrared quantities
such as $\Gamma_d$ is at best an expansion in $\lambda / g^2$, and we
know of no reason why the expansion for $\Gamma_d$ should be better
behaved than the expansion for $ \langle \phi^2 \rangle $.
At tree level, the phase transition is second order, and $\Gamma_d$
is large just below $T_{\rm c}$ \cite{ArnoldMcLerran}.  
At one loop, the phase transition becomes first order and 
$\Gamma_d$ is exponentially suppressed up to $T_{\rm c}$.  
To see how well perturbation theory converges,
we would like a two loop calculation.  Our experience from the
calculation of $ \langle \phi^2 \rangle $ is that two loop
corrections may not be small.

Unfortunately, extending the perturbative calculation of $\Gamma_d$ to
two loops is immensely harder than it is for $\langle \phi^2
\rangle$.  Both are background field calculations, but the background
field in the effective potential calculation is spatially uniform, and
the diagrams can be computed using Fourier techniques.  The sphaleron
background, in contrast, lacks translation symmetry, and is only known
numerically.  The one loop calculation demands summing the eigenvalues of
numerically determined fluctuation eigenmodes in this background.  At
two loops one needs to compute mutual interactions between the
fluctuations, which will involve double and triple sums over the
eigenmodes of overlap integrals; and the interaction Hamiltonian
depends on the numerically determined sphaleron background.  
In our opinion it is
actually easier to make a direct nonperturbative calculation of
$\Gamma_d$ than to perform this two loop calculation.

In the symmetric phase we can now determine $\Gamma_d$ with fair
reliability, using
real time techniques
\cite{slavepaper,AmbKras2,particles}\footnote{Recently it has been
pointed out \protect{\cite{Bodek_log}}, correctly, that in the
parametric limit of small $\alpha_w$ there are logarithmic corrections
to the functional form assumed for fitting in
\protect{\cite{particles}}.  The numerical value of the coefficient of
the log can be computed in a simple effective theory
\protect{\cite{Bodek_log}}, and the result is that the coefficient of
the log is small \protect{\cite{inprep}}; so while there is an
uncontrolled systematic in the most recent results for the symmetric
phase rate \protect{\cite{particles}}, the error caused is relatively
small.}, 
but this does not work in the broken phase, because the rate is too 
small for a transition to occur in a reasonable amount of Hamiltonian
time.  But if we knew what we meant by the horizontal axis in Figure
\ref{Fig1}, then we could use nonperturbative lattice tools to find the
diffusion constant.  Both the probability to be within $\epsilon / 2$ of
the peak, and the instantaneous mean value of $| dN_{\rm CS} / dt |$, are
thermodynamic properties, and while the probability to be near the peak
is exponentially small, we could measure it using multicanonical
Monte-Carlo techniques.  That would give $\Gamma_d$ up to an order unity
dynamical prefactor, reflecting the fact that when the system crosses
the peak, it does not necessarily get stuck in the new phase, but may
turn around and re-cross immediately.  In the
broken phase we expect this dynamical prefactor to be on order 1;
we will estimate it in the conclusion.

Our approach in this paper is precisely to find a way to make sense of
Fig. \ref{Fig1} in a nonperturbative lattice study of the dimensionally
reduced electroweak theory, and to construct the figure and find
the probability to be at the peak by multicanonical Monte-Carlo
techniques.  We also need to measure $ \langle | dN_{\rm CS} / dt | \rangle$
at the peak, meaning the value averaged over the ensemble restricted to
configurations with $N_{\rm CS} = 1/2$.  
This paper will present the crux of the technique,
with the essential ideas; but we will postpone
some technical details to a longer sequel \cite{nextpaper}.  
The outline of the paper is
as follows.  First we construct a suitable definition
of $N_{\rm CS}$.  Then we use it to perform a multicanonical Monte-Carlo 
determination  of $\Gamma_d$.  We compare our results at three values of 
$x \equiv \lambda / g^2$ to results of 
a perturbative calculation of
$\Gamma_d$ made using the two loop effective potential.  We determine 
that, in the minimal standard model, or any theory where heavy
degrees of freedom can be integrated out until it looks like the minimal
standard model \cite{MSSMDR}, the sphaleron
bound at the physical value of the Weinberg angle is $x \equiv \lambda /
g^2 \simeq 0.036$.

\section{Definition of $N_{\rm CS}$}

We will treat the thermodynamics of the hot standard model in the
dimensional reduction approximation, so the configuration space and
probability distribution are the same as thermal, classical Yang-Mills
theory with certain Higgs mass counterterms.  We use notation and field
normalizations appropriate for the 3+1 dimensional classical field
theory and will determine the instantaneous value of 
$ | dN_{\rm CS} / dt | $ in the classical theory context.  But we need
an appropriate definition of $N_{\rm CS}$.

Two features are essential to the definition of $N_{\rm CS}$:
\begin{enumerate}
	\item $N_{\rm CS}$ is the integral of the total derivative 
	      $(g^2 / 8 \pi^2) E_i^a B_i^a$; and
	\item $N_{\rm CS}$ is an integer for a vacuum configuration, 
	      $B_i^a = 0$.
\end{enumerate}
To determine the rate $\Gamma_d$ by constructing Fig. \ref{Fig1} we only
need to define $N_{\rm CS}$ modulo 1; so we can set $N_{\rm CS}=0$ for
all vacua, and we may hope for a gauge invariant definition.

To measure $N_{\rm CS}$ of some three dimensional gauge field
configuration we then find a path from that configuration to a vacuum
and integrate $E_i^a B_i^a$ along that path.  The most natural path to
choose is the cooling path; following \cite{AmbKras2} we define a
cooling or dissipative time $\tau$, under which gauge fields evolve by
dissipative dynamics,
\begin{equation}
\frac{dA_i^a(x,\tau)}{d\tau} = - \frac{\partial H_{\rm YM}(A(\tau))}
	{\partial A_i^a(x,\tau)} \, , \qquad A_i^a(x,0) = A_i^a(x) \, .
\end{equation}
We write in continuum notation for clarity; the covariant lattice
implementation is in \cite{AmbKras2}, and Hetrick and de Forcrand have
used a roughly equivalent procedure to solve the Gribov gauge fixing
problem \cite{Hetrick}.  We cool only the gauge fields, under the
Yang-Mills Hamiltonian, since $N_{\rm CS}$ should be a function of gauge
fields alone; not cooling the Higgs fields also avoids some technical
problems.  

In a finite volume, the gauge fields will cool towards a vacuum off a
measure zero subspace of configurations which cool towards saddlepoints.
Hence we can define $N_{\rm CS}$ as
\begin{equation}
N_{\rm CS} = \frac{g^2}{8 \pi^2} \int_{\tau_0}^{\infty} d \tau
	\int d^3 x E_i^a(x,\tau) B_i^a(x,\tau) \, ,
\label{trueNCS}
\end{equation}
where $E_i^a(\tau)$ is the field strength in the $i,\tau$ direction.  
The standard lattice implementation of $E_i^a B_i^a$ \cite{AmbKras} is
not a total derivative \cite{Moore1}, but it is to good approximation
for smooth fields, by which we mean fields with almost no UV excitation.
Cooling removes UV excitations much more efficiently than IR ones--in a
linear theory the amplitude of a mode decays as $\exp( - \tau k^2)$--so
the shortcomings of the definition of $E_i^a B_i^a$, and regulation
sensitivity more generally, go away at a very small value of $\tau$, 
and are not a problem.
Also note that the cooling procedure and the definition of 
$E_i^a B_i^a$ are gauge invariant, and so is our definition of $N_{\rm
CS}$. 

We cut off the small $\tau$ part of the integration to remove UV
contributions to $N_{\rm CS}$ which are unrelated to winding number
change.  In the continuum abelian theory, where
\begin{equation}
N_{\rm CS} = \frac{ g^2 }{32 \pi^2} \int d^3 x \epsilon_{ijk} 
	 F_{ij} A_k \, ,
\label{naiveNCS}
\end{equation}
the mean square value of $N_{\rm CS}$ we would get if we used 
$\tau_0 = 0$ is
\begin{equation}
\langle N^2_{\rm CS} \rangle = \frac{g^4}{1024 \pi^4} \int d^3 x d^3 y
	\epsilon_{ijk} \epsilon_{lmn} \langle F_{ij}(x) A_k(x) F_{lm}(y)
	A_n ( y ) \rangle \, .
\end{equation}
Using Wick's theorem and the momentum representation of the propagator,
in Feynman gauge, this becomes
\begin{eqnarray}
\langle N^2_{\rm CS} \rangle & = & \frac{g^4}{1024 \pi^4} \int d^3 x d^3 y
	\int \frac{d^3 p d^3 q}{(2 \pi)^6} e^{i (p+q) \cdot (x-y)}
	\epsilon_{ijk} \epsilon_{lmn} \left( \frac{4 p_i p_l 
	T^2 \delta_{jm} \delta_{kn} }{p^2 q^2} 
	+ \frac{4 p_i q_l T^2 \delta_{jn} 
	\delta_{km} }{p^2 q^2} \right)  \nonumber \\
& = & \frac{g^4 T^2 V}{64 \pi^4} \int \frac{d^3 p}
	{ (2 \pi)^3 } \frac{p^2}{(p^2)^2} \, .
\label{UVdiv}
\end{eqnarray}
so $N_{\rm CS}$ will be Gaussian distributed with a linearly divergent 
variance.  On the
lattice, the UV divergence will be cut off by the lattice scale; the
coefficient was found by Amjorn and Krasnitz \cite{AmbKras} and is
$\langle N^2_{\rm CS} \rangle = (1.44 \times 10^{-5}) g^4 V T^2 / a$.  

SU(2) theory will have similar UV contributions, since
in the UV it looks like three copies of the abelian theory.  This UV
contribution has nothing do with topology, since it appears already in
the abelian theory.  If we used $\tau_0 = 0$ then the measured value of
$N_{\rm CS}$ would be an ``interesting'' IR piece plus this
nontopological UV piece.  The probability distribution of $N_{\rm CS}$
would be the convolution of the probability distributions for the two,
and if the noise distribution were broad, this would spread $N^{\rm IR}_{\rm
CS} \simeq 0$ configurations out to dominate the sample at $N_{\rm
CS} = 1/2$, destroying the peak in Fig. \ref{Fig1}.  Only if the noise
distribution is narrow will the probability distribution be undistorted.

At finite $\tau_0$, and including the mass for the gauge fields which
appears in the broken phase, the UV contribution to $N_{\rm CS}$,
Eq. (\ref{UVdiv}), becomes approximately
\begin{equation}
\langle N_{\rm CS}^2 \rangle \simeq \frac{g^4 T^2 V}{64 \pi^4} 
	\int \frac{d^3 p} { (2 \pi)^3 } \frac{p^2 e^{-4 p^2 \tau_0}}
	{(p^2 + m_W^2)^2} \, .
\end{equation}
(Note that $\tau$ has units of length squared, not length.)
Choosing $\tau_0 \ge 1/m_W^2$ controls the noise sufficiently, while
leaving $N_{\rm CS}$ with the feature which is essential to our
endeavor, namely that a configuration which is very nearly a sphaleron
(meaning that it cools to very close to the ``sphaleron'' saddlepoint,
but then slips off and finds a vacuum) will have $N_{\rm CS}$ very close
to $\pm 1/2$.  We have checked that the part of the cooling path from 
(almost) the sphaleron saddlepoint to the vacuum gives $(g^2 / 8 \pi^2)
\int E_i^a B_i^a d^3 x d \tau$ of almost exactly $\pm 1/2$.

This definition of $N_{\rm CS}$ has a somewhat arbitrary parameter
$\tau_0$, so we cannot give a direct physical interpretation to the 
probability distribution of $N_{\rm CS}$.  Also,
the probability distribution of $N_{\rm CS}$ will depend on $\tau_0$.
Consider a configuration which is ``fairly close to'' a sphaleron but
which is already rolling away from the saddlepoint at cooling time
$\tau_0$.  If we increase $\tau_0$, it will roll further out of the
saddlepoint before we start to integrate $E \cdot B$, and if we decrease
$\tau_0$ then it will be closer to the saddlepoint.  Hence, choosing a
larger $\tau_0$ will spread out the determined values of $N_{\rm CS}$ of
similar, nearly sphaleron configurations
more widely, diluting the probability distribution near $N_{\rm CS} =
1/2$ (and concentrating the probability distribution near $N_{\rm CS} =
0$).  This reduces the population of configurations with $N_{\rm
CS}$ within $\epsilon$ of $1/2$.  However, spreading out the
configurations increases $dN_{\rm
CS} / dt$, evaluated on an ensemble of configurations with $N = 1/2$, 
by ``almost'' the same factor, and so the rate $\Gamma_d$ ``almost'' does
not depend on $\tau_0$.  By ``almost'' we mean that
this statement is strictly true when we can neglect the contribution to
$|dN/dt|$ arising from residual UV fluctuations.
The more we cool, the larger the
contribution to $dN/dt$ from motion in the unstable direction and the
smaller the contributions from the stable directions, so the truer this
becomes.  In a complete calculation including the ``dynamical prefactor''
this remaining $\tau_0$ dependence is also accounted for
\cite{nextpaper}.  For this work we will use a value of $\tau_0$ where we
can directly check that fluctuations about the sphaleron are largely
absent for configurations with $N_{\rm CS} = 1/2$.

We make two further approximations to make measuring
$N_{\rm CS}$ numerically practical.  The cooling is very numerically
expensive, but after cooling by $\tau > a^2$ all UV information on the
lattice has been destroyed, and cooling on such a fine lattice is
redundant.  We set up a coarsened lattice of all even sites and
continue the cooling on it.  We have compared the value of $N_{\rm CS}$
obtained in this way to the value without coarsening, and aside from a
slight and measurable renormalization, the difference is a tiny amount
of noise.  We also extrapolate the large $\tau$ exponential tail of
$\int E \cdot B d \tau$  to avoid having to run $\tau$ to arbitrarily
large times.  

\section{Multicanonical measurement}

Now that we have a definition of $N_{\rm CS}$ (modulo 1), we
want to measure the probability to have a particular value, that is
\begin{equation}
P(x) \equiv \frac{1}{Z \delta x} \int {\cal D} \Phi {\cal D} A 
	e^{- \beta H (\Phi,A)} \Theta ( N_{\rm CS} - x ) 
	\Theta ( x + \delta x - N_{\rm CS} ) \, ,
\label{defofP}
\end{equation}
where $Z$ is the same integral without the $\Theta$ functions, which
select out the range of $N_{\rm CS}$ 
between $x$ and $x + \delta x$.   We are
making the dimensional reduction approximation including the integration
over the $A_0$ field, though we do include the U(1) hypercharge field at
the physical value of $\Theta_W$ the Weinberg angle, using the noncompact
lattice implementation.

To determine $P(x)$ accurately even where it is small, we perform a
multicanonical Monte-Carlo calculation.  
That is, we rewrite Eq. (\ref{defofP}) as
\begin{equation}
P(x) \equiv \frac{1}{Z \delta x} \int \left( {\cal D} \Phi {\cal D} A 
	e^{- \beta H (\Phi,A)} e^{f(N_{\rm CS})} \right) e^{-f(N_{\rm CS})} 
	\Theta ( N_{\rm CS} - x ) 
	\Theta ( x + \delta x - N_{\rm CS} ) \, ,
\label{MCMC}
\end{equation}
where $f(x)$ is a function chosen by bootstrapping 
to be approximately equal to $- \ln P(x)$.  
We do the integral by finding a Markovian process which samples
configuration space with weight $e^{- \beta H (\Phi,A)} e^{f(N_{\rm
CS})} {\cal D} \Phi {\cal D} A$, and replacing the integral with a
sum over a sample generated by that Markovian process.  Since $N_{\rm
CS}$ is a complicated function of the configuration, this is easiest
done by finding a Markovian process which samples with canonical weight
and adding a Metropolis accept reject step to account for $e^f$.

We generate the canonical Markovian process with the algorithm from
\cite{KLRSresults}, extending it to include the noncompact U(1) subgroup,  
and {\rm reducing} the efficiency of
the gauge field updates, to bring the Metropolis accept rate 
up to about $50 \%$.  To measure $| dN_{\rm CS} / dt|$, we
augment the system with
Gaussian ``momentum'' degrees of freedom, pick momenta from the thermal
distribution, and evolve all fields forward under Hamilton's equations
(implemented by leapfrog) for a (very short) time.  We determine $dN_{\rm
CS} / dt$ from the difference between $N_{\rm CS}$ of the initial and
final configurations.  This differs from a Hamiltonian evolution of the
classical system because we do not enforce Gauss' law; but since
$B$ is transverse and Gauss' law only
affects the longitudinal $E$ field, the instantaneous value of 
$E \cdot B$ should not be affected.

\begin{figure}[t]
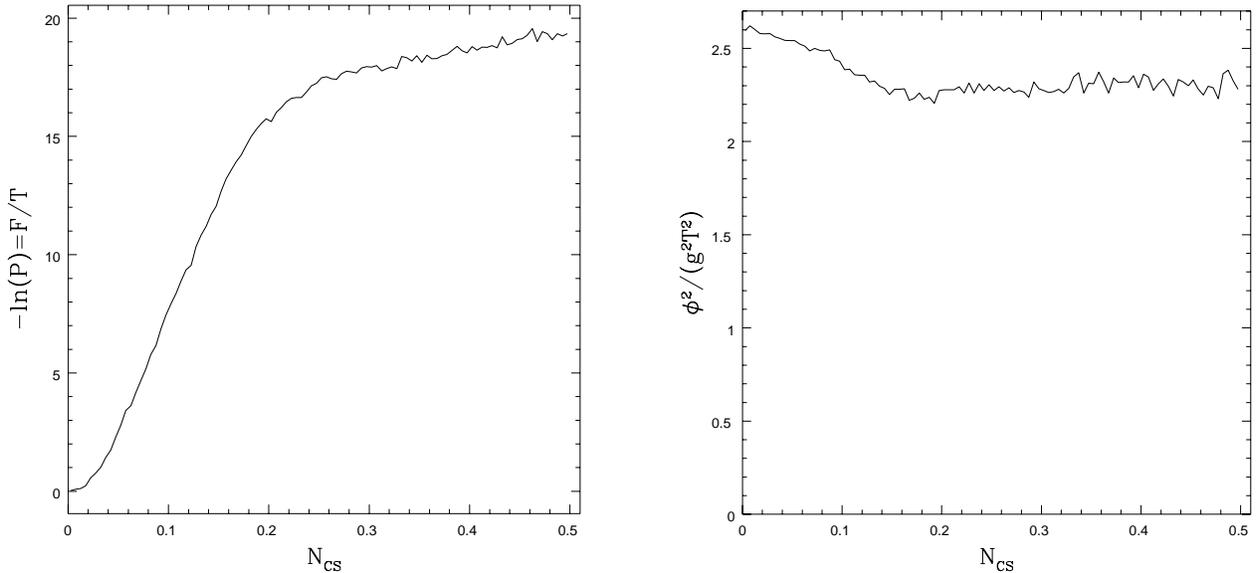

\centerline{\mbox{\psfig{file=fig3a.epsi,width=3in}} \hspace{0.4in}
\mbox{\psfig{file=fig3b.epsi,width=3in}}}
\caption{ \label{Fig3} Free energy (left) and $( \phi^2_{\rm
broken} - \phi^2_{\rm symm} ) / ( g^2 T^2 ) $ as functions of
$N_{\rm CS}$ at $x \equiv \lambda / g^2 = 0.039$, in a $(16/g^2 T)^3$
volume.  The plot of $\phi^2$
is a check that the volume used was large enough to prevent the
sphaleron from causing a transition to the symmetric phase.}
\end{figure}

We have measured $\Gamma_d$ for three values of scalar self-coupling, $x
\equiv (\lambda / g^2) = 0.047$, $0.039$, and $0.033$.  
We used a $40^3$ box with periodic boundary conditions and a lattice 
spacing of $a = 2 / (5 g^2 T)$ ($\beta_{\rm L} = 10$) for the
larger values of $x$ and $a = 1/(3 g^2 T)$ ($\beta_{\rm L} = 12$) 
for $x=0.033$.  We determine the equilibrium temperature as in 
\cite{Fodor}.  The lattice action is $O(a)$ improved using the relations
in \cite{Oapaper,Oa2}, so finite lattice spacing systematic errors 
will be smaller than statistical errors.  
The very large volume is necessary to prevent 
the tails of the sphaleron from meeting because of the periodic boundary
conditions.  It is also necessary because the Higgs field 
condensate has a zero at
the core of a sphaleron, so the sphaleron looks somewhat like a symmetric
phase bubble; the large volume prevents the sphaleron from
stimulating a transition to the symmetric phase.

We plot the numerically determined
free energy and $\langle \phi^2 \rangle$ versus $N_{\rm CS}$ 
for the $x=0.039$ data in Fig. \ref{Fig3}.  The value of $\langle |
dN_{\rm CS} / dt | \rangle$ at $N_{\rm CS} \simeq 0.5$ was $(0.22 \pm 0.05)
a^{-1} = (0.55 \pm 0.13) g^2 T$.  Recall that both this value, and free
energy distribution, depend somewhat on our choice of $\tau_0$, which
was $\tau_0 = 22.5 a^2 = 3.6 / g^4 T^2$.  With less cooling, the free
energy would rise more evenly and would be less flat on top; so the
shape of the free energy curve should not be overinterpreted.  The
determined rate $\Gamma_d$ should be independent of $\tau_0$, though.  

We present the results in Table \ref{Table1}.  They should be
compared to the symmetric phase rate, which, using 
$g^2 = 0.40$, is $\Gamma_d = (29 \pm 6) \alpha_W^5 T^4 = 
\exp(-13.9 \pm 0.2) T^4$ \cite{particles}\footnote{Again, this result
misses a logarithmic correction which is computable but not computed
\protect{\cite{Bodek_log}}.}.
As expected, the rate is orders of magnitude smaller in the broken phase
than in the symmetric phase.

\begin{table}
\centerline{\mbox{\begin{tabular}{|cc|c|c|c|}\hline
 & & $x = 0.047$ & $x = 0.039$ & $ x = 0.033 $ \\ \hline
&   $\phi(T_{\rm c}) / gT_{\rm c}$  &  1.360  
	& 1.568  &  1.789  \\ \cline{2-5}
``2 loop'' & $B \equiv g E_{\rm sph} / 4 \pi \phi$ & 1.643  &  1.633  
	&  1.626  \\ \cline{2-5}
perturbative& $E_{\rm sph} / T_{\rm c}$  &  28.08  &  32.20  
	&  36.55  \\ \cline{2-5}
 & $- \ln (\Gamma_d T_{\rm c}^{-4})$ &  22.27  &  25.39  
	&  28.82  \\ \cline{2-5}  
& $ - d \ln (\Gamma_d T^{-4}_{\rm c} ) / dy $  
	&  860  &  920  &  1000  \\ \hline
nonperturbative  &  $\phi(T_{\rm c}) /gT_{\rm c}$  &  $1.38 \pm 0.02$  
	&  $1.60 \pm 0.01$  & $1.82 \pm 0.03$   \\ \cline{2-5}
 &  $- \ln ( \Gamma_d T_{\rm c}^{-4} )$  &  $24.7 \pm 0.4$  &  
	$28.3 \pm 0.4$  & $31.2 \pm 0.6$  \\ \hline
mixed & $- \ln (\Gamma_d T_{\rm c}^{-4})$ & $22.6 \pm 0.3$  
	& $25.9 \pm 0.2$    &  $29.3 \pm 0.5$   \\ \hline
\end{tabular}}}
\caption{ \label{Table1}
Perturbation theory versus nonperturbative $\Gamma_d$.
Appearances of $T_{\rm c}^{-4}$ are really $(2.5 g^2 T_{\rm c})^{-4}$,
and $d/dy$ means $g_w^4 d/d(m_{\rm H}^2/T^2)$.
The error bars for the nonperturbative $\phi(T_{\rm c})$ are dominated by 
statistical errors
in the determined value of $T_{\rm c}$; errors in the nonperturbative
value of $\Gamma$ are statistical errors from the Monte-Carlo.  The
``mixed'' results use the two loop value of $B$ but the
nonperturbative value of $\phi(T_{\rm c})$.}
\end{table}

\section{Comparison to perturbation theory, erasure bound}

Next we compare the rate to a perturbative estimate.  One loop
perturbation theory gives \cite{ArnoldMcLerran}\footnote{The 
definition of $\Gamma$ used in \protect{\cite{ArnoldMcLerran,Carson}} 
is the response rate to a chemical potential,  which is
half the diffusion rate \protect{\cite{KhlebShap}}; 
so Eq. (\protect{\ref{Gammapert}}) differs by a factor of 2
from the expressions in those references.}  
\begin{equation}
\Gamma_d = 4 T^4 \frac{\omega_{-}}{g \phi} \left( \frac{\alpha_W}{4
	\pi} \right)^4 \left( \frac{4 \pi \phi}{g T} \right)^7 
	{\cal N}_{\rm tr} {\cal NV}_{\rm rot} \kappa e^{- \beta E_{\rm
	sph}} \, .
\label{Gammapert}
\end{equation}
Here $\phi$ is
the broken phase Higgs condensate expectation value, $\omega_{-}$
is the unstable frequency of the sphaleron, ${\cal N}_{\rm tr} 
{\cal NV}_{\rm rot}$ are zero mode factors, $\kappa$ is the one loop
fluctuation determinant, and $E_{\rm sph}$ is the energy of the
Klinkhamer Manton sphaleron, using the tree level Hamiltonian.  
For small scalar self-coupling, $- T \ln \kappa$ equals the energy due 
to the one loop effective potential term, plus a modest correction
\cite{Baacke}.  We can guess that the dominant two loop corrections to
Eq. (\ref{Gammapert}) are absorbed by including the two loop
effective potential terms in the Hamiltonian.  So it seems reasonable to
estimate the sphaleron rate by Eq. (\ref{Gammapert}), but setting
$\kappa = 1$ and solving for the sphaleron energy using the two loop
effective potential at the equilibrium temperature.  One should also
solve for the zero modes and $\omega_{-} / \phi$ 
at this value, but they are very
weak functions of the effective potential \cite{Carson}.  We use the
values from \cite{Carson} at $x = (\lambda / g^2) = 0.04$ for these, but
solve for the sphaleron energy, $E_{\rm sph} = 4 \pi B \phi / g$,
numerically, using the two loop effective potential at $T_{\rm c}$.  We 
use the two loop potential presented in \cite{Hebecker}, without pieces 
from longitudinal gauge bosons (assumed integrated out). We also drop
two loop terms proportional to $\lambda g^2$ or $\lambda^2$, because the
perturbative determination of $\phi$ is an expansion in $\lambda /
g^2$, and such terms contribute at the same or higher order as unknown 3
loop terms.  (Including those terms moves $\phi$ closer to the
nonperturbative value.)  We compare the results to the numerically 
determined values in Table \ref{Table1}.  The ``two loop'' analytic 
sphaleron rate is about $\exp(2.5)$ times faster than the numerically 
determined nonperturbative rate.  The difference is more than can be 
explained by the difference in $\phi$.

We should compare the sphaleron rate to the limit set by
requiring that baryon number not be erased.  The rate at which
sphalerons degrade baryon number is \cite{ArnoldMcLerran}
\begin{equation}
\frac{1}{N_{\rm B}} \frac{d N_{\rm B}}{dt} = 
	- \frac{13 N_{\rm F}}{4} \Gamma_d T^{-3} \, ,
\end{equation}
where $N_{\rm F}=3$ is the number of families, and the numerical 
factor $13 N_{\rm F}/4$ would be smaller in theories, such as
supersymmetry,  in which additional degrees of freedom can store 
baryon number.\footnote{Again, there is a factor of 2 difference from 
the reference because they write in terms of the response to a chemical
potential, which is half the diffusion constant.}
Integrating from the end of the phase transition to the
present day,
\begin{equation}
\ln ( N_{\rm B} / N_{\rm B}(T_{\rm c}) ) 
	= - \frac{13 N_{\rm F}}{4} \int_{t_0}^\infty \Gamma_d(T(t)) 
	T^{-3}(t) dt \, ,
\end{equation}
where we have shown the dependence of $\Gamma_d$ on $T$ and of $T$ on
$t$.  

Now $\ln \Gamma_d$ is very
sensitive to $\sqrt{\langle \phi^2 \rangle}$, and hence to
$T$; so we can
approximate $\ln \Gamma_d(T) \simeq \ln \Gamma_d(T_{\rm c}) + 
(T-T_{\rm c}) (d \ln\Gamma_d/dT)|_{T=T_{\rm c}}$, 
and perform the integral:
\begin{equation}
\ln ( - \ln ( N_{\rm B} / N_{\rm B}(T_{\rm c}) ) ) = \ln(39/4) + 
	\ln \left( \frac{\Gamma_d(T_{\rm c})}{T^{4}} \right) 
	- \ln \left( \left. - \frac{ d \ln \Gamma_d(T(t)) }{T dt} 
	\right|_{T=T_{\rm c}} \right) \, .
\label{doublelog}
\end{equation}
By the chain rule,
\begin{equation}
\frac{d \ln \Gamma_d}{T dt} = \frac{d \ln \Gamma_d}{d y} \frac{dy}{dT}
	\frac{d \ln T}{dt} \, ,
\end{equation}
where $y = m_{\rm H}^{2}(T) / (g^4 T^2)$ is the dimensionless thermal
Higgs mass squared.  

Two of these are easy.  We get $dy/dT$ from the 1 loop correction to
$m_{\rm H}^2$ \cite{Hebecker},
\begin{equation}
\frac{dy}{dT} \simeq \frac{8 \lambda + 4 g_{\rm y}^2 + 
	g^2 (3 + \tan^2 \Theta_W)}{8 g^4 T} \, ,
\end{equation}
and we get $d \ln T/dt$ from the Friedmann equation in a radiation
dominated universe,
\begin{equation}
\frac{1}{4 t^2} = H^2 = \frac{8 \pi G}{3} 
	\frac{ \pi^2 g_*}{30} T^4  \quad \Rightarrow \quad 
	\frac{d \ln T}{dt} = - \sqrt{ \frac{4 \pi^3 g_*}{45} }
	\frac{T^2}{m_{\rm pl}}
	\, ,
\end{equation}
where $g_*$ is the number of radiative degees of freedom in 
the universe ($g_* = 106.75$ in the minimal standard model) 
and $m_{\rm pl} \simeq 1.22 \times 10^{19}$GeV is the Planck
mass.  Finally, we determine $d\ln \Gamma_d / dy$ perturbatively, by 
varying $y$ slightly from the equilibrium value and recomputing the 
two loop sphaleron rate.  The dependence is quite strong.  
We include it in the table.

The most widely cited discussion of baryon number erasure after the
phase transition makes the approximation that the baryon number
violation rate after the phase transition is constant for about one
Hubble time \cite{Shapwrong}.  In fact,
because $\Gamma_d$ depends very strongly on $y$, which in turn depends
strongly on $T$, most baryon number erasure occurs in the first $10^{-3}$
Hubble times after the phase transition.  Hence the initial rate of baryon
number violation,
$\Gamma_d(T_{\rm c})$, which prevents washout is $10^3$ times 
larger than assumed in \cite{Shapwrong}, leading to a weaker
bound on $\Gamma_d(T_{\rm c})$, roughly
\begin{equation}
- \ln (\Gamma_d(T_{\rm c}) T_{\rm c}^{-4} ) 
	> 30.4 - \ln ( T_{\rm c} / 100{\rm GeV} )\, .
\label{thebound}
\end{equation}
The values of $g_*$ and $dy/dT$ are both larger in supersymmetric 
extensions of the standard model, by on order a factor of 2; so the 
bound, Eq. (\ref{thebound}), is weaker in those models 
by about 1.  Also note that, because
Eq. (\ref{doublelog}) is for the double log of $N_{\rm B} / N_{\rm
B}(T_{\rm c})$, failing to meet the bound by 1 means the baryon number is
diminished by $\exp(\exp(1)) \simeq 15$, and failure by 2 reduces baryon
number by $\exp(\exp(2)) \simeq 1600$; so the bound is quite sharp.

Interpolating between the values of $x$ where we have measured, and
including the estimate discussed in the conclusion for the dynamical 
prefactor, we get a bound of about $x=0.036$ in the standard model and
$x=0.038$ in the MSSM.

\section{Conclusion}

Our nonperturbative results show that the bound on $x \equiv \lambda /
g^2$ is about $0.036$ in the standard 
model.  This result will be corrected only slightly in any
theory where all non-standard model degrees of freedom are heavy enough
to integrate out accurately.  This should apply for instance to the
minimal supersymmetric standard model unless the lightest stop is quite
light \cite{MSSMDR}.  The bound weakens
somewhat if the latent heat of the phase transition is not sufficient to
reheat the universe to $T_{\rm c}$; this should be checked in the MSSM.  

If additional degrees of freedom are light, then they can affect the
thermodynamics of the sphaleron (aside from renormalizing couplings).
The calculation should then be redone, including those degrees of
freedom.  However, we have shown that if a two loop perturbative
calculation gets the value of $\phi$ about right, then its value for
$\Gamma_d$ is reasonably close.  We do not expect light 
stop squarks to endanger this conclusion; while their contribution to
the effective potential may be important, they only contribute to the
scalar wave function at two loops, and they do not interact directly
with the gauge fields at all.  Hence, they should change the sphaleron
bound mainly by changing the size of the Higgs condensate.
We can paraphrase the results of this work as setting a 
bound on $\phi$ of roughly $\phi \ge 1.7 gT$, and this should
be approximately the same in extensions with a light stop.

Finally we should comment on the approximation that successive sphaleron
crossings will be in uncorrelated directions.  Arnold, Son, and Yaffe
have recently criticized this assumption \cite{ArnoldYaffe}, on the
grounds that it ignores dynamical effects related to hard thermal loops.
Their picture has recently been borne out in the symmetric phase
\cite{particles}.  In the present context, we would expect the system to
cross the sphaleron barrier multiple times before relaxing to a vacuum
configuration if the plasma frequency $\omega_{\rm pl}$ is much greater
than the unstable mode frequency $\omega_{-}$.  It should be clear why;
when $\omega_{\rm pl} \gg \omega_{-}$, 
the electric field, and hence $E_i^a B_i^a$, will 
oscillate in sign, driving $N_{\rm CS}$ back and
forth, on a time scale short compared to the time scale at which it
``falls off'' the sphaleron.  On the other hand, if $\omega_{\rm pl} < 
\omega_{-}$, then we expect no such effect.  A crude, parametrically
correct estimate is that the rate we determined will be 
corrected by a factor of
\begin{equation}
\frac{1}{1 + \omega_{\rm pl}^2 / \omega_{-}^2} = 
	\left[ 1 + \left( \frac{11 g^2 T^2}{18} \right) 
	( 0.47 g^2 \phi^2 )^{-1} \right]^{-1} \, .
\end{equation}
Here we use the standard model value for $\omega_{\rm pl}^2$ and the
value of $\omega_{-}^2$ computed in \cite{Carson} for $\lambda \simeq
0.03 g^2$.  Numerically, using $g^2 \simeq 0.4$, this estimate reduces the
rate by a factor of 2 for $x = 0.033$, changing
the bound on $x = \lambda / g^2$ by  $\simeq 0.0015$.
The correction in the symmetric phase is larger, because the natural 
length scale of a winding number changing process is larger; if we
estimate that scale as $g^2 T/2$, for instance, the correction would be
$\sim 7$.  It is possible to use the techniques of
\cite{particles} to measure the prefactor numerically \cite{nextpaper}, 
but we will not do so here.

\section*{Acknowledgments}

I would like to thank Jim Cline, Jim Hetrick, Kari Rummukainen,
and Alex Krasnitz for discussions and correspondence.  I also thank 
Berndt M\"{u}ller and the North Carolina Supercomputing Center.

\end{document}